\documentclass[sn-basic]{sn-jnl}

\usepackage{graphicx}%
\usepackage{multirow}%
\usepackage{amsmath,amssymb,amsfonts}%
\usepackage{amsthm}%
\usepackage{mathrsfs}%
\usepackage[title]{appendix}%
\usepackage{xcolor}%
\usepackage{textcomp}%
\usepackage{manyfoot}%
\usepackage{booktabs}%
\usepackage{algorithm}%
\usepackage{algorithmicx}%
\usepackage{algpseudocode}%
\usepackage{listings}%

\newcommand{\bs}{{\boldsymbol \sigma}}

\newcommand{\btt}{{\boldsymbol t}}

\newcommand{\be}{{\boldsymbol e}}

\newcommand{\bv}{{\boldsymbol v}}

\newcommand{\bF}{{\boldsymbol F}}

\newcommand{\F}{{\cal F}}

\newcommand{\bD}{\boldsymbol  D}
\newcommand{\bW}{\boldsymbol  W}

\newcommand{\bni}{{\boldsymbol n}}
\newcommand{\bI}{{\boldsymbol I}}

\newcommand{\D}{ {\cal D}}

\def\R{\mathbb{R}}

\def\phi{\varphi}

\def\incep{\left\{\begin{array}{ll} }
 \def\termin{\end{array}\right. }

\def\la2{\lambda^2}

\begin{document}

\title[Article Title]{Accounting for localized deformation: a simple computation of true stress in  micropillar compression experiments}


\author[1]{\fnm{Jalal} \sur{Smiri}}\email{jalal.smiri@edu.sorbonne-paris-nord.fr}
\equalcont{These authors contributed equally to this work.}
\author[1]{\fnm{O\u{g}uz Umut} \sur{Salman}}\email{oguzumut.salman@cnrs.fr}
\equalcont{These authors contributed equally to this work.}
\author[1]{\fnm{Matteo} \sur{Ghidelli}}\email{matteo.ghidelli@lspm.cnrs.fr}
\equalcont{These authors contributed equally to this work.}
\author*[1,2]{\fnm{Ioan R.} \sur{Ionescu}}\email{ioan.r.ionescu@gmail.com}
\equalcont{These authors contributed equally to this work.}

\affil*[1]{\orgdiv{LSPM}, \orgname{University Sorbonne Paris Nord}, \orgaddress{\street{av. J.P Clement}, \city{Villetaneuse}, \postcode{93430}, \country{France}}}

\affil[2]{\orgdiv{IMAR}, \orgname{Romanian Academy}, \orgaddress{\street{str. Grivitei}, \city{Bucharest}, \postcode{10587},  \country{Romania}}}


\abstract{
Background: Compression experiments are widely used to study the mechanical properties of materials at micro- and nanoscale. However, the conventional engineering stress measurement method used in these experiments neglects to account for the alterations in the material's shape during loading.  This can lead to inaccurate stress values and potentially misleading conclusions about the material's mechanical behavior, especially in the case of localized deformation.

Objective: Our goal is to calculate true stress in cases of localized plastic deformation from standard experimental data (displacement-force curve, aspect ratio, shear band angle and elastic strain limit). 

Methods:  We use a simple mechanical-geometrical approach based on reasonable physical assumptions to get analytic formulas of true stress and eliminating the need for finite element computations. 
Furthermore, in numerical simulations of pillar compression, the formula-based true stress demonstrates strong alignment with the theoretical true stress.

Results :  We propose analytic formulas  for calculating true stress in cases of localized plastic deformation commonly encountered in experimental settings for  a single band  oriented in arbitrary directions with respect to the vertical axis of the pillar.

Conclusions : The true stress computed with the proposed formulas provides a more precise interpretation of experimental results and can serve as a valuable and simple tool in material design and characterization.
}



\date{}
\maketitle

\section{Introduction}\label{sec1}

Compression experiments conducted on pillars have proven to be a valuable method for analyzing the mechanical behavior of materials at the micro- and nano-scales. This approach involves  the fabrication of micro-pillars (often with focused ion beam (FIB) techniques) followed by an  uni-axial compression to study its mechanical response   in  a deformation  process under  displacement or load control. This method has been particularly useful for investigating the onset and evolution of plastic deformation in materials,  by  exploring the local deformation mechanism (when compression test are carried out in situ SEM), see for instance \cite{Uchic2004-ax,Greer2005-ak,Ng2008-ii,Frick2008-dr,Kiener2011-sn,Friedman2012-ie,Zhang2017-cg,Salman2021-ts,Zhang2020-ax,Salman2021-sn,Cui2021-st}. Specifically, micro-pillar compression experiments have revealed numerous new phenomena, including the transition from wild-to-mild plasticity \cite{Zhang2017-cg}, pristine-to-pristine plastic deformation \cite{Wang2012-aw}, the "smaller is stronger" effect \cite{Lee2009-as}, size- and shape-dependent flow stresses \cite{Uchic2004-ax,Maas2012-ib,Zhang2017-ff} and, microstructural control of plastic flow~\cite{Rizzardi2022-iz},  among others.

During such compression experiments, the material can undergo significant plastic deformation, which can manifest in either homogeneous deformation or slip/kink bands~\cite{Hagihara2016-fa,Mayer2016-tp,Basak2019-mw,Chen2020-ay,Nandam2021-hi,Weiss2021-db,Nizolek2021-gj,Marano2021-bh,Zhang2022-if}.  Homogeneous deformation occurs when the material undergoes uniform deformation throughout its structure, while slip/kink bands result from localized deformation that can form along some preferred orientation~\cite{Inamura2019-he,Gan2020-yk}. The resulting engineering strain-stress curve is related to a displacement-force experimental recording, but in order to accurately characterize the material's mechanical behavior, it is necessary to determine the Eulerian (true) stress that is exerted within the deformation zone.  It is especially crucial to be able to  accurately interpret the mechanical properties of engineered or designed materials using various methods to assess whether desired enhancements have been achieved~\cite{Wu2021-ex,Ghidelli2021-us}.  The significance of using true stress in assessing mechanical responses has been discussed in prior studies related to  the mechanical behavior of  metallic glass~\cite{Han2008-zr,Wang2017-dh}. However, of particular importance is the fact that,  to the best of our knowledge,  there is currently no established method to calculate the required load-bearing area to evaluate true stress, during plastic localization mechanisms.

In this context, the aim of this study is to derive simple formulas for calculating true stress in cases involving slip/kink band formation during mechanical loading while avoiding the need for lengthy and complex finite element computations that deal with large deformations of crystals.  Specifically, we consider a  localization,  observed frequently in experiments as single band oriented in arbitrary directions with respect to the vertical axis  of the pillar, for which  we derive a formula and employ it to assess the reliability of previous experimental results.  We have to mention here that the proposed formula is completely geometric. Contrary to   Finite Elements (FEà computations, it does not need any  material modeling   setup, hence it  could be very useful  in choice of  the  constitutive law.
\begin{figure*}[ht!]
	\includegraphics[scale=.35]{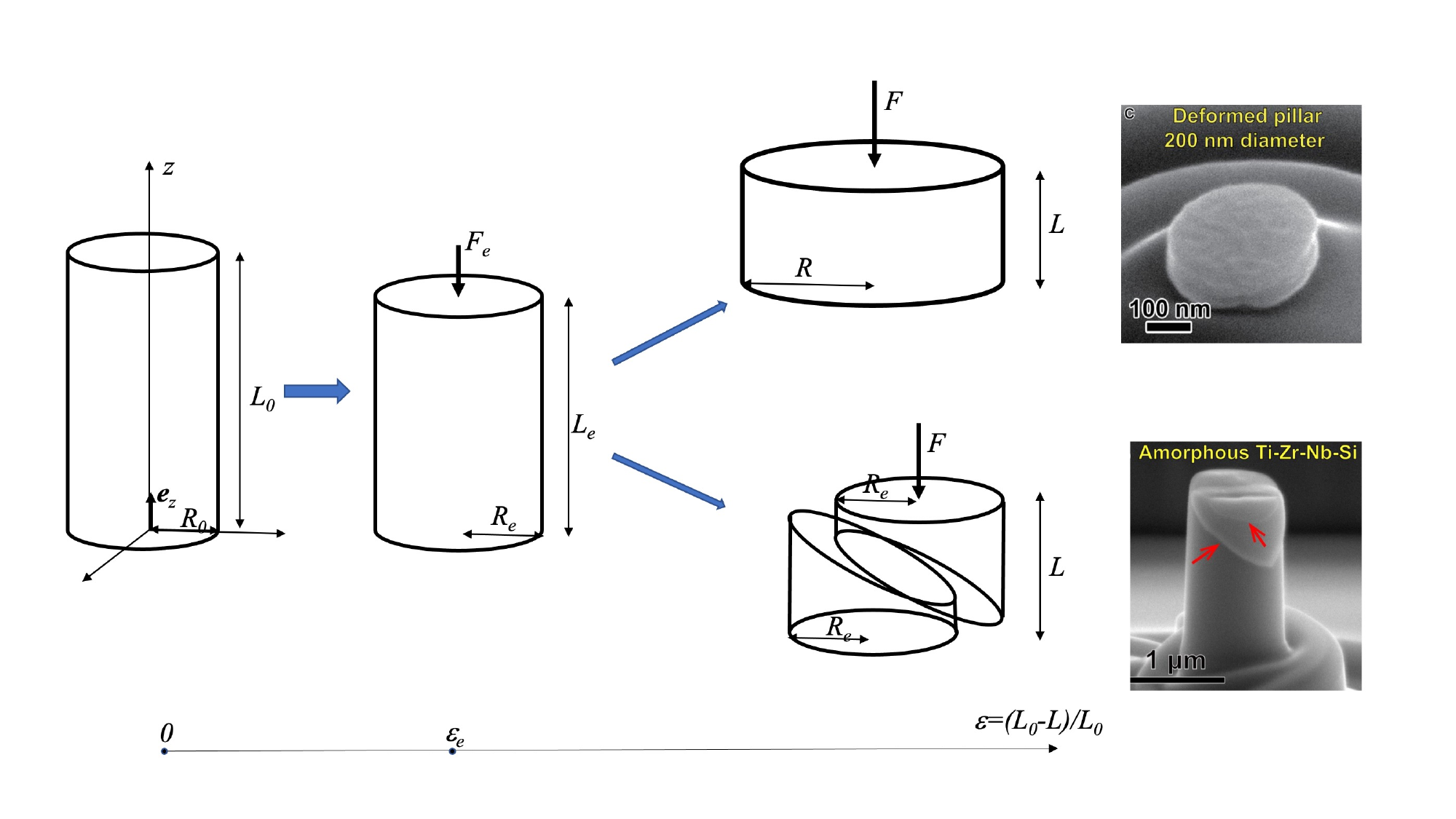}
	\caption{Schematic representation of the nano or micro-pillar deformation. The linear elastic regime, $\epsilon^{eng} < \epsilon_e$, is followed by one of the two types of plastic flow. Up:  homogeneous deformation, Bottom:  shear/kick band deformation (experimental illustration taken from \cite{Wu2021-ex}).}
	\label{GenView}
\end{figure*}

\section{Simple modeling of pillars'  deformation}

After the initial loading process, which is associated with small-strain linear elastic behavior, the  pillars undergo significant plastic deformation, making the elastic deformations negligible in comparison to the plastic ones. From these plastic deformation processes, two distinct scenarios emerge: homogeneous and slip/kink band, as illustrated schematically in Figure \ref{GenView} and detailed subsequently. The Cauchy stress tensors corresponding to these two deformation mechanisms exhibit different patterns. In either scenario, the primary challenge is to determine the true stress $\sigma^{true}$ within the uniaxial Cauchy stress tensor $\bs=-\sigma^{true}\be_z\otimes\be_z$, where $\be_i$ represents the elements of the orthonormal basis of the three-dimensional Euclidean vector space, acting on the active area $A_u$.

To be more specific, let $R_0$ and $L_0$ represent the initial (Lagrangian) radius and height of the cylindrical pillar, respectively, while $R$ and $L$ denote the current (Eulerian) dimensions during deformation, as shown in Fig. \ref{GenView}. Let $\epsilon^{eng}=(L_0-L)/L_0$ denote the overall engineering strain. Let $\bF=-F\be_z$ represent the force applied to the top of the pillar during deformation, where $F=\sigma^{true}A_u$, and let $\sigma^{eng}$denote the nominal (engineering) stress, i.e., $F=\sigma^{eng}A_0$, with $A_0=\pi R_0^2$ is the original cross-sectional area.

We assume knowledge of the initial pillar shape, specifically the aspect ratio $f_0=L_0/2R_0$, and have access to the engineering strain-stress curve, denoted as the function $\epsilon^{eng}\to \sigma^{eng}(\epsilon^{eng})$. The primary objective of this paper is to derive a simple formula for estimating the engineering strain-true stress curve, represented as $\epsilon^{eng}\to \sigma^{true}(\epsilon^{eng})$.
\begin{figure*}[ht!]
	\centering
	\includegraphics[scale=.15]{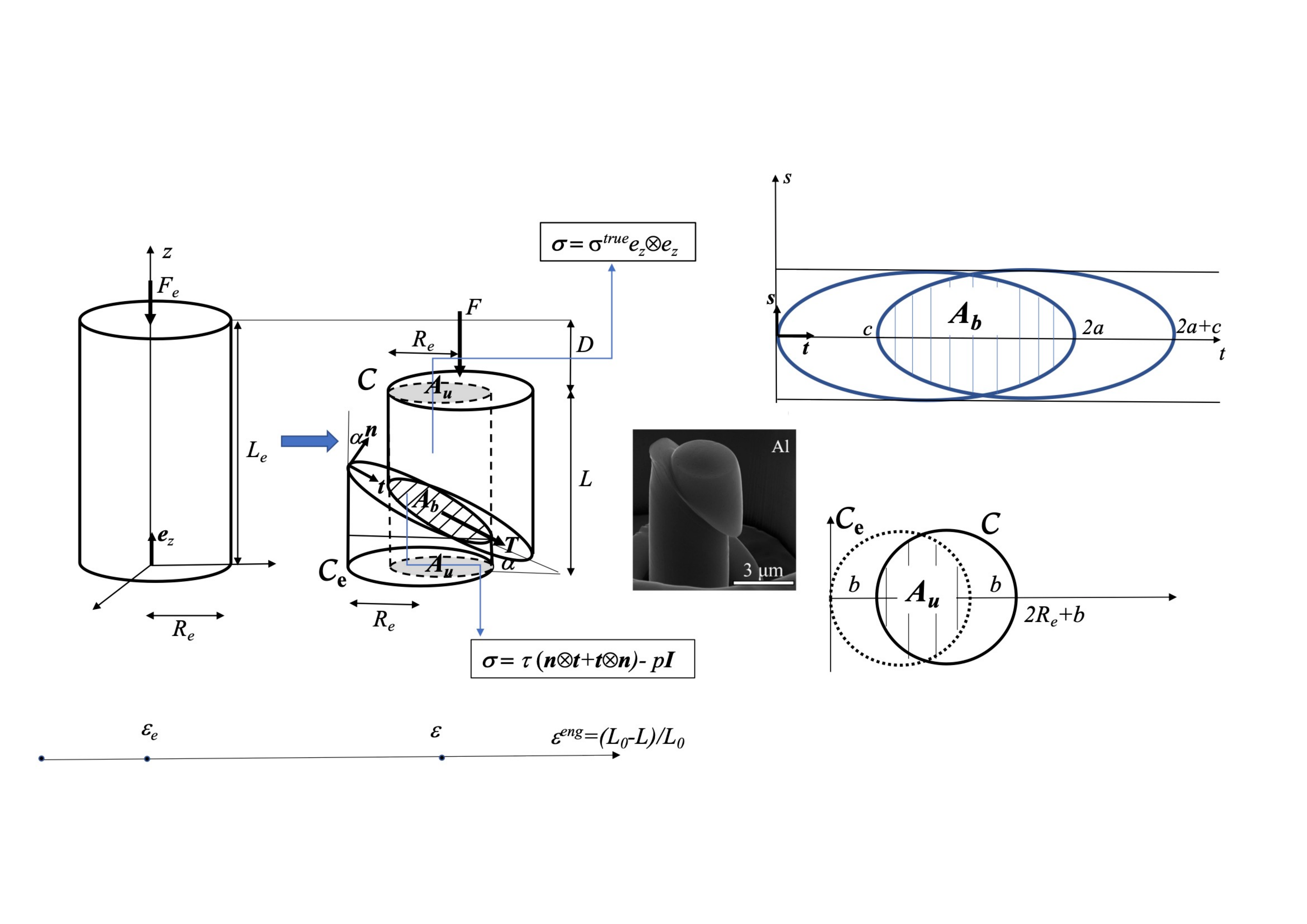}
	\vspace{-1.cm}
	\caption{(Left): Schematic representation of localized plastic deformation following the elastic stage with the Cauchy stress tensor acting in different regions of the pillar. (Right, top): The plan of the shear band with its area $A_b$ between two ellipses representing the upper and lower sections of the pillar. (Right, bottom) : Its projection on the horizontal plane (experimental image taken from \cite{Weiss2021-db}). }
	\label{Localiz}
\end{figure*}

\subsection{Elastic  deformation}
For   $\epsilon^{eng}<\epsilon_e$ (or equivalently for $\sigma^{eng}<\sigma^{eng}_e$) the pillar exhibits  a linear elastic behavior. Here,  $\epsilon_e$ and $\sigma^{eng}_e=\sigma^{eng}(\epsilon_e)$ represent the strain and stress limits of elasticity, which   can be  easily identified in each stress-strain (or force-displacement)  curve.
Since the elastic strain limit  $\epsilon_e$, is usually small (less that $3\%$)  the elastic linear theory can be accepted as a good approximation. If  the deformed shape is also a cylinder  and the stress is uniaxial throughout the pillar then   $A_u=A=\pi R^2, F=\sigma^{true}\pi R^2$ and $\sigma^{eng}R_0^2=\sigma^{true} R^2$. For anisotropic materials, such as monocrystals, during the elastic phase, the pillar is no longer a perfect cylinder. However, since the deformation is small, the deviation from a cylindrical shape can be neglected.

Following the Hooke's law  the volumetric strain  $\epsilon_V=(V_0-V)/V_0=(L_0R_0^2-LR^2)/L_0R_0^2=(1-(1-\epsilon^{eng})(R/R_0)^2)$ is related to  the true stress through the  compressibility modulus $K$  by $\sigma^{true}=3K\epsilon_V$.    
We get a  second order algebraic equation for $(R_0/R)^2$ to deduce that  $(R_0/R)^2=(1-\sqrt{1-4\delta(1-\epsilon^{eng})})/2\delta$, where $\delta=\sigma^{eng}/3K$.  Finally we have  
$$ \sigma^{true}=\frac{3K-\sqrt{9K^2-12K\sigma^{eng}(1-\epsilon^{eng})}}{2},  \quad \mbox{for} \; \epsilon^{eng}\leq\epsilon_e, $$
which for small values of $\delta_e=\sigma^{eng}_e/3K$   and $\epsilon_e$,  gives the  well-known  formula for true stress:
\begin{equation}\label{TrueElast}
\sigma^{true} = \sigma^{eng}(1-\epsilon^{eng}), \quad  \quad \mbox{for} \; \epsilon^{eng}\leq\epsilon_e.
\end{equation}
Note that since   $1-\epsilon^{eng}\approx 1$ for $\epsilon^{eng}<\epsilon_e$ the difference between the true and engineering stress in not significant, and we can conclude that $	\sigma^{true} \approx  \sigma^{eng}$  during the elastic phase. 

In most strain-stress curves, the elastic strain limit $\epsilon_e$ is usually easy to extract by considering the end of the linear behavior. In the following, $\epsilon_e$ will be considered as obtained from the experimental results, but this assumption will not imply any further considerations (i.e., $\epsilon_e$) about the hardening or softening plastic behavior of the pillar.


\subsection{Homogeneous uni-axial stress}
For larger deformations,   $\epsilon^{eng}>\epsilon_e$, in the first scenario the  deformation is homogeneous and the stress throughout the entire pillar is assumed to be uni-axial, given by $\bs=-\sigma^{true}\be_z\otimes\be_z$, where $A_u=A=\pi R^2$. For materials modeled by a pressure-independent plasticity law, plastic deformation is isochoric and the volume is preserved.    If the  deformed shape remains a cylinder and  we neglect  the  further elastic deformation of the volume then we get    $V=\pi R^2L=V_e=\pi R_e^2L_e$, in which   $R_e$ and $L_e$ are, respectively, the radius and length of the pillar at the end of  elastic phase (i.e. fixed  during post-elastic deformation). After some algebra we find  
$$  \sigma^{true}=\frac{\sigma^{true}_e}{\sigma^{eng}_e(1-\epsilon_e)}\sigma^{eng}(1-\epsilon^{eng}), \quad \mbox{for} \; \epsilon^{eng} >\epsilon_e.
$$
Since for small values of $\delta_e=\sigma^{eng}_e/3K$ we have  $\sigma^{true}_e=\sigma^{eng}_e(1-\epsilon_e)$  we get the well known  formula 
\begin{equation}\label{TrueHom}
\sigma^{true} = \sigma^{eng}(1-\epsilon^{eng}), \quad \mbox{for} \; \epsilon^{eng} >\epsilon_e.
\end{equation}
However, for large values of $\epsilon^{eng}$, using the nominal stress $\sigma^{eng}$ instead of the Cauchy stress $\sigma^{true}$ can significantly alter the behavior of the stress-strain diagram, giving a false impression of overall hardening-like behavior.

Due to the boundary conditions on the top and bottom, the above assumption concerning the cylindrical shape of the deformed sample is not always valid. Indeed, barreling or bulging phenomena could occur, and the sample shape is given by two diameters (the middle one and the top/bottom one). If the minimum between the top/bottom and center radii, denoted by $R_m(\epsilon^{eng})$, could be measured during the experiment, then true stress can be directly computed through $\sigma^{true}=F/(\pi R_m^2(\epsilon^{eng}))$, and we do not need the above formula.

%
%
%

\subsection{Slip/kink band  plastic deformation}

In the second scenario, for  $\epsilon^{eng}>\epsilon_e$,  deformation is localized in a narrow zone between two parallel planes with a normal vector $\bni$,  determined by the angle $\alpha$ with respect to the vertical axis $\be_z$, see Fig. \ref{Localiz}.  {\color{blue} The Cauchy stress tensor acting in the shear band is given by $\bs=\tau(\bni\otimes\btt+\btt\otimes\bni) -p\bI$, where $\btt$ is the slip direction,  $\tau$ is the shear stress and, $\bI$ is the identity matrix,   while   in two cylindrical regions above and below the shear band it is assumed to be uniaxial, i.e., $\bs=\sigma^{true}\be_z\otimes\be_z$ (see Fig. \ref{Localiz}).   This assumption is a schematic representation of the stress distribution in the pillar  with three non-vanishing  uniform  stress zones which allows  analytic computations.  Even if the stress distribution  is expected to be much more complicate this assumption seems to be  globally verified in FE computations (see for instance Fig \ref{TruNum} bottom). }

From the above assumption we can deduce that the expression for  shear stress $\tau$, acting in the shear band, is proportional to the true stress:
\begin{equation}\label{tau}
\tau=\frac{1}{2}\sin(2\alpha)\sigma^{true}.	
\end{equation}
It should also be noted that if the true stress $\sigma^{true}$ is known, then equation (\ref{tau}) enables the calculation of the shear stress $\tau$ as a function of the shear plastic strain $\gamma_p$, which is proportional to the plastic axial engineering strain and can be expressed as
\[
\gamma_p=\frac{L_e-L}{\cos(\alpha)H_b}=\frac{(\epsilon^{eng}-\epsilon_e)L_0}{\cos(\alpha)H_b},
\]
in which $H_b$ represents the thickness of the shear band. {\color{blue} The diagram of shear stress $\tau$ versus shear plastic strain $\gamma_p$ is a very important tool in any discussion about the choice of the plastic model to be considered, both in crystal plasticity and for amorphous materials.}

Let us now compute the area $A_u$  between the two disks  (or equivalently $A_b=A_u/\cos(\alpha)$  the area between the   two ellipses) ${\cal C}_e$ and $\cal{C}$ corresponding to the projection on the basal plane of the two cylinders (see Fig. \ref{Localiz}). One of the circles is translated by a distance of $b=D\cot(\alpha)$, where $D=L_e-L$ is the vertical displacement of the upper pillar region. After some simple computations, one can find that the area $A_u$ between the two regions is given by 
$$ A_u=R_e^2\left[\pi -\frac{b}{R_e}\sqrt{1-\frac{b^2}{4R_e^2}}-2\arcsin\left(\frac{b}{2R_e}\right)\right].$$
Denoting by $f_0=L_0/2R_0$ the initial shape number and by $f_e=L_e/2R_e=(1-\epsilon_e)^{3/2}f_0$ the shape number at the end of the elastic phase, and by $\epsilon_*^{eng}=(Le-L)/L_e=(\epsilon^{eng}-\epsilon_e)/(1-\epsilon_e)$ the engineering (plastic) deformation with respect to the configuration at the end of the elastic phase, we get 
\begin{equation}
\label{Au}
A_u=R_e^2 \Phi(\epsilon_*^{eng}f_e\cot(\alpha)) ,
\end{equation}
where we have denoted by 
$$\Phi(s)=\pi-2s\sqrt{1-s^2}-2\arcsin(s).$$
Taking  into account that $R_0^2/R_e^2=(1-\sqrt{1-4\delta_e(1-\epsilon_e)})/2\delta_e$ we can deduce that:
\begin{equation}
\label{TrueExact}
\sigma^{true}=\frac{\sigma^{eng}\pi (1-\sqrt{1-4\delta_e(1-\epsilon_e)})}{2\delta_e\Phi((\epsilon^{eng}-\epsilon_e)f_e\cot(\alpha)/(1-\epsilon_e))}, \quad \mbox{for}\;  \epsilon^{eng}> \epsilon_e.
\end{equation}

Bearing in mind that  $\epsilon_*^{eng}f_e=(\epsilon^{eng}-\epsilon_e)f_0\sqrt{1-\epsilon_e}$ for small values of $\epsilon_e$ and $\delta_e$, we have $R_0^2/R_e^2 \approx 1-\epsilon_e$  we can deduce  a simplified formula for the true stress:

\begin{equation}
\label{Truefinal}
\sigma^{true}=\frac{\pi (1-\epsilon_e)\sigma^{eng}}{\Phi(  (\epsilon^{eng}-\epsilon_e)f_0\sqrt{1-\epsilon_e}\cot(\alpha))} , \quad \mbox{for} \; \epsilon^{eng}> \epsilon_e
\end{equation}

Note that to use the simplified formula we only need to know the elastic limit $\epsilon_e$, the shear band angle $\alpha$  and the initial aspect ratio $f_0$. However, the exact formula given in Eq. \ref{TrueExact} requires also   $\delta_e$: the ratio between engineering stress $\sigma^{eng}_e$ at the end of the elastic phase and the bulk modulus  $K$.

In contrast to the homogeneous deformation scenario, here the true stress is larger than the engineering stress. Therefore, in many strain-stress diagrams, the plateau or softening of the engineering stress should be viewed as a hardening of the true stress.

\section{Comparison with 2-D  FE computations}

In the above model, the shear band thickness is assumed to be small relative to the specimen length (i.e., $H_b/L_0 \ll 1$), a condition verified in many situations. For experiments where this assumption is not verified, Eqs. (\ref{TrueExact}) and (\ref{Truefinal}) need to be revisited. Moreover, due to the Lagrangian description of large plastic deformations in the shear band, finite element simulations must contend with severe distortion of elements. Consequently, their results exhibit a shear band thickness that is unrealistically large, and computations are halted at intermediate strains (less than 20\%). Consequently, a conclusive comparison between the above formulas and Lagrangian FE computations could not be made.

\begin{figure}[hbt!]  
	\begin{centering}
		\includegraphics[scale=0.14]{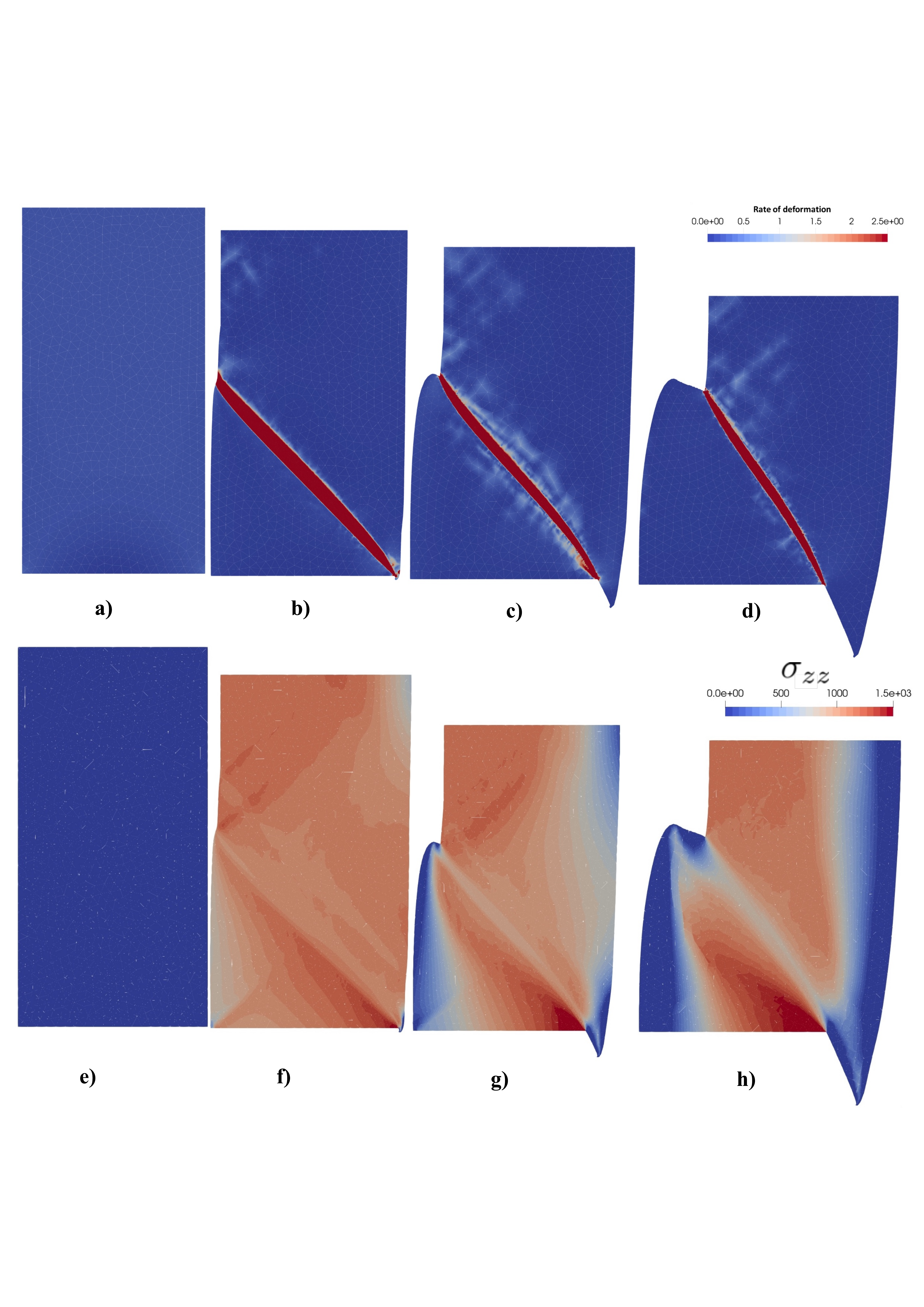} 
		\vspace{-2cm}
		\caption{2D   Eulerian FE  computation of the shear band  localization in an elastic-perfectly plastic pillar with the plastic strain rate $\dot{\varepsilon^p}$ {\color{blue}(up) and the Cauchy stress $\sigma_{zz}$ (bottom)} in  color scale. Evolution from  for different values of $\epsilon^{eng}$: 0\% in a), 5.5\% in b), 11\% in c) and 22\% in d). }
		\label{LocPlane}
	\end{centering}		
\end{figure}

For this reason, we will use here an Eulerian approach in the FE computations, capable of describing thin shear bands. We will select an elastic perfectly plastic material (see the Appendix for the constitutive equation) with the Von-Mises yield limit $\sigma_Y$, such that the theoretical shear stress in the shear band is known ($\tau_{th}=\sigma_Y/\sqrt{3}$). From Equation (\ref{tau}), we can compute the theoretical true stress $\sigma_{th}^{true}=2\sigma_Y/(\sqrt{3}\sin(2\alpha))$ for $\epsilon^{eng}> \epsilon_e$. Then, $\sigma_{th}^{true}$ can be compared with the true stress $\sigma^{true}$ computed from Equations (\ref{TrueExact}) or (\ref{Truefinal}) and the FE computations of the engineering stress $\sigma^{eng}$.

The ALE approaches of the shear bands, which require re-meshing at each time step, are computationally very expensive. Therefore, we have performed only 2-D computations here. For the two-dimensional case, the geometric true stress formula (Equation (\ref{Truefinal})) reads:

\begin{equation}
\label{Truefinal2D}
\sigma^{true} = \frac{(1-\epsilon_e)\sigma^{eng}}{1- (\epsilon^{eng}-\epsilon_e)f_0(1-\epsilon_e)\cot(\alpha)}, \quad \text{for} \; \epsilon^{eng}> \epsilon_e.
\end{equation}

In Figure \ref{LocPlane}, we have plotted the evolution of the shear band in an elastic-perfectly plastic pillar of initial shape number $f_0=2$, with the following material constants: $E=38235 \text{ MPa}$, $\nu=0.34$, $\sigma_Y=1000 \text{ MPa}$. We remark that the ALE computations of the Eulerian model were able to handle a thin shear band. The angle of the shear band is, as expected, $\alpha=45°$ at the beginning, but we observe a slight variation at the end of the deformation process. {\color{blue}  At the bottom of Fig. \ref{LocPlane}, the distribution of the Cauchy stress $\sigma_{zz}$ is plotted.   One can remark that this distribution has,   globally,  a  good agreement with the schematic stress distribution  plotted in Figure \ref{Localiz}.  } 

\begin{figure}[hbt!]  
	\begin{centering}
		\includegraphics[scale=0.55]{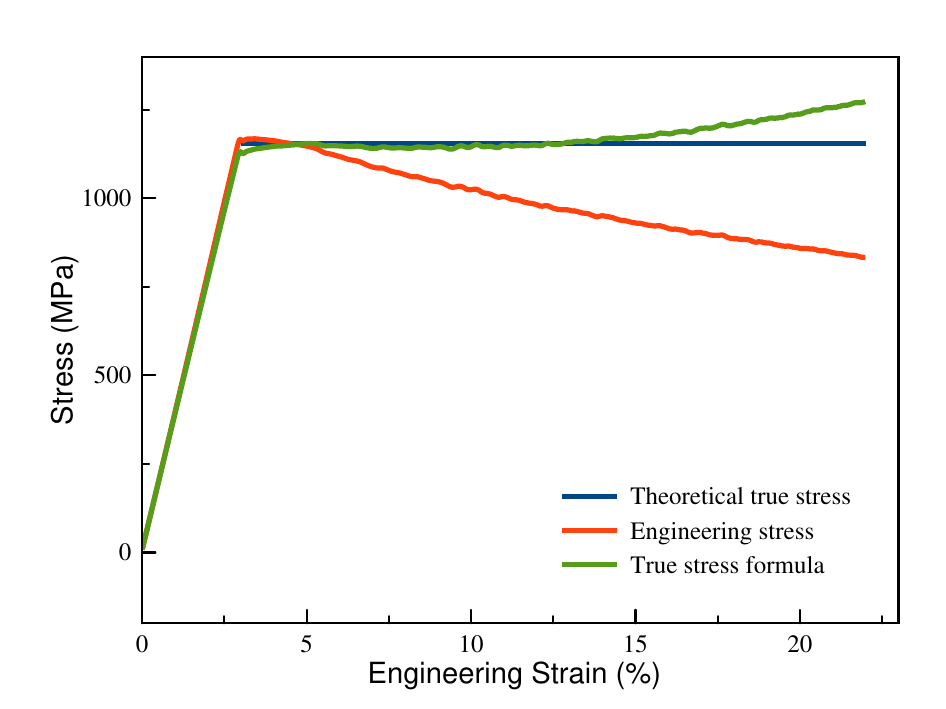} 
		\vspace{-0.25cm}
		\caption{FE computations  with an Eulerian elastic perfectly plastic model  of the engineering  stress $\sigma^{eng}$ (in orange).   In green the true stress $\sigma^{true}$ computed  with the formula (\ref{Truefinal2D})  and in blue  the theoretical true stress  $\sigma_{th}^{true}$ that we expect from the model. }
		\label{TruNum}
	\end{centering}		
\end{figure}

In Figure \ref{TruNum}, we have plotted the engineering stress $\sigma^{eng}$ (in orange) computed from the resultant force on the pillar's top of the FE simulations. From this curve, we obtain the elastic strain to be $\epsilon_e=0.031$. In blue, we have plotted the theoretical true stress $\sigma_{th}^{true}$ that we expect from the model, and in green, the true stress $\sigma^{true}$ computed with Equation (\ref{Truefinal2D}). We observe that the formula-based true stress closely aligns with the theoretical true stress up to $\epsilon^{eng}<18\%$. After $\epsilon^{eng}=18\%$, the true stress overestimates the theoretical one. This is due to the variation of the shear band angle $\alpha$, which is larger at the end of the deformation process.

\section{True stress computation  and re-interpretation of  the stress-strain  curves}
In this section, we want to illustrate how the formulas deduced in the previous section alter some experimental engineering strain-stress curves reported in the literature.

\begin{figure}[hbt!]  
	\begin{centering}
		\includegraphics[scale=0.1]{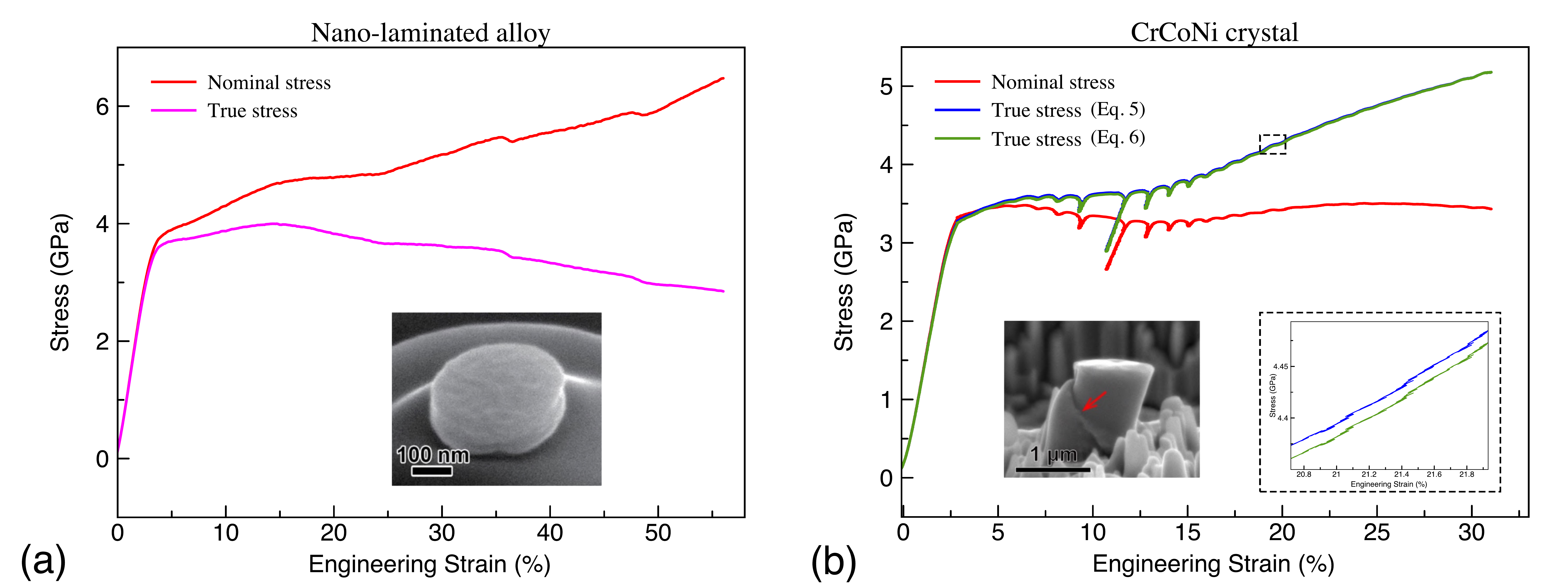} 
		\caption{Engineering strain vs. nominal stress (red) and true stress (blue) curves. Nominal stress values are taken from \cite{Wu2021-ex}) and the calculated true stress curves are calculated using the formulas derived here.  (a)  homogeneous deformation of a crystal-glass nano-laminated alloy sample calculated with the classical formulas (\ref{TrueElast}), (\ref{TrueHom}); (b) single shear band deformation of a CrCoNi crystal calculated with formulas (\ref{TrueElast}), (\ref{TrueExact}) (blue) and simplified formula (\ref{Truefinal}) (green).  Insets show the final state of the pillar at the end of the loading.}
		\label{Crystal1}
	\end{centering}		
\end{figure}

As an example, we will reconsider the strain-stress behavior of a crystal-glass symbiotic alloy investigated in \cite{Wu2021-ex}. In this study, the authors plotted the engineering strain-stress curves to characterize the crystal-glass nano-laminated alloy sample's mechanical properties compared with its crystalline and amorphous counterparts. Interestingly, the authors found that the nano-laminated crystal-glass alloy appeared to be tougher than its individual components when analyzing the engineering stress data (see Fig. 4(a) in \cite{Wu2021-ex}). However, when we calculate  the engineering strain vs. true stress curves for two cases:
(i) the crystal-glass nano-laminated alloy undergoing homogeneous deformation using the conventional formula provided in \eqref{TrueHom}; (ii) the CrCoNi crystal experiencing slip band deformation using (\ref{TrueExact}) with $\alpha=\pi/4$, $f_0=$1, $\epsilon_e=$2.5\% and  $\sigma^{eng}_e=$3.23 GPa (or the simplified formula given in (\ref{Truefinal}), we observe a contradictory outcome in both cases. Specifically, the nano-laminated alloy exhibits softening, as shown in Fig. \ref{Crystal1}(a), while the CrCoNi crystal demonstrates hardening, as depicted in Fig. \ref{Crystal1}(b). This finding emphasizes the significance of taking into account the current deformation state of the material, even in the absence of a strong localization.

{We have to notice that, for the nano-laminated crystal-glass alloy, we deal with a bulging deformation, and one could expect that the true stress obtained through \eqref{TrueHom} is slightly underestimated. Indeed, from the final shape of the pillar, we observe that the central radius $R_m$ is less than 3\% of the volume-preserving radius $R$, hence the underestimation of true stress is less than 6\% at the end of the deformation. This is negligible with respect to the difference between engineering and true stress, plotted in Fig. \ref{Crystal1}(a), which is on the order of 100\%, and cannot change the above conclusions.

	Lastly, we emphasize that when comparing the true stress curves obtained using  (\ref{TrueExact}) and the simplified formulas provided in  (\ref{Truefinal}), as shown in Fig. \ref{Crystal1}(b), we observed minimal differences, even at high strains. 

	\section{Conclusion}
	In conclusion, our study provides formulas for calculating true stress in cases where slip/kink bands form during mechanical loading in compression experiments on pillars. These formulas are  simple and need only the engineering stress data,  some geometric data (aspect ratio), and some mechanical data (elastic limit) which are easy  to get from the experimental results.  {\color{blue}  For slip/kink band plastic deformation the shear stress  $\tau$ acting in the band can be recovered  from the uniaxial true stress $\sigma^{true}$. The diagram shear plastic strain- shear  stress    is then the main information on the mechanical behavior  of the material that can obtained from a pillar  experiment with a shear band localization.   }

	Of course, our formula does not consider bending and torsion, which cannot be easily traced by simple geometrical arguments. However, in our opinion, their effects will be of second-order.  In a 2D finite element numerical simulation of Eulerian elasto-plastic pillar compression, we compared the formula-based true stress with the theoretical true stress and we found a  good agreement.

	A more precise alternative to these simple formulas could be very long and difficult FE computations involving large deformations. Moreover, the FE computations needs to know in advance  the material type (mono-crystal, poly-crystal, isotropic),  its behavior  (hardening, softening, etc) and the  constants that characterize the material.  {\color{blue}This means that when localization occurs, the proposed geometric formula gives the possibility  to experimenters to interpret the data in a simple manner without any preconceived notions about the choice of model.  }

	However, using these formulas, we re-evaluated the robustness of previous experimental results and found that considering the current deformation state of engineered materials can be important for accurately interpreting their mechanical behavior at small scales.  To be more precise, our analysis revealed that, in some cases, the true stress led to conclusions that were exactly opposite to those found using the engineering stress, while in other cases, the difference is mainly quantitative and the overall trend is similar. To conclude, our work provides a valuable tool for accurately interpreting the mechanical behavior of materials under compressive loads and for drawing appropriate conclusions based on the true stress values.

\bmhead{Acknowledgments}

O.U.S. is supported by the grants ANR-18-CE42-0017-03, ANR-19-CE08-0010-01, ANR-20-CE91-0010, M.G. is supported by the grant ANR-21-CE08-0003-01 and I.I. acknowledges the support of the Romanian Ministry of Research (PN-III-P4-PCE-2021-0921 within PNCDI III).


\section*{Declarations}

The authors  disclose any interests that are directly or indirectly related to the present paper.  It  has obtained approval for publication from all co-authors and does not involve any  human or animals participants.



\section{Annex:  Elastic perfectly plastic Eulerian model }

The  movement (flow) in the Eulerian description is given by the velocity field, denoted $\bv(t,\cdot):\D_t\to \R^d$ (here $\D_t$ is the Eulerian domain occupied by the elasto-plastic body at time $t$). The rate of deformation and the spin rate are denoted by $\bD=\bD(\bv)=(\nabla\bv+\nabla^T\bv)/2$ and by $\bW=\bW(\bv)=(\nabla\bv-\nabla^T\bv)/2$, respectively while  the Cauchy stress tensor is  $\bs(t,\cdot):\D_t\to\R^{d\times d}_S$.  To describe the elasto-plastic Eulerian  model  (see for instance  \cite{Belytschko2014Nonlinear}), we consider  the additive decomposition of the rate deformation tensor into the elastic $\bD^e$ and plastic rates $\bD^p$ of deformation 
$$ \bD=\bD^e+\bD^p.$$ 
For the  elastic range  we considered the generalization of Hooke's law  written in terms of  the Jaumann rate of the Cauchy stress tensor   ${\bs}^\nabla=\dot{\bs}-\bW\bs-\bs\bW$ (here  $\dot{\bs}=\partial_t\bs+\bv\cdot\nabla \bs$ is the total derivative)  given by 
$$ {\bs}^\nabla (t)=\lambda \text{trace}(\bD^e)\bI+2\mu\bD^e, \quad \mbox{in} \: \D_t,$$
where  $\lambda, \mu$ are the Lam\'e  elastic coefficients. The plastic rate of deformation is related to the  Cauchy stress tensor through the flow rule associated to the classical  Von-Mises yield criterion with no hardening (perfectly plastic material). To be more precise, let $\F(\bs)=\sigma_{eq}-\sigma_Y$ be the yield function, with $\sigma_Y$ the yield limit and $\sigma_{eq}=\sqrt{\frac{3}{2}}\vert \bs^D\vert$ the Von-Mises stress ($\bs^D=\bs-\frac{1}{3}\text{trace}(\bs)\bI$ is the stress deviator).  If we denote the accumulated plastic strain by $\varepsilon^p$ (given through the differential equation $\dot{\varepsilon^p}=\sqrt{\frac{3}{2}}\vert \bD^p\vert$), then  the flow rule and the   loading-unloading conditions read 
$$\bD^p=\frac{\dot{\varepsilon^p}}{\sigma_{eq}}\bs^D, \quad \dot{\varepsilon^p}\geq 0, \quad \F(\bs)\leq 0, \quad \dot{\varepsilon^p} \F(\bs)= 0.$$


\end{document}